\documentclass[pdflatex,sn-mathphys-num]{sn-jnl}

\usepackage{hyperref}
\usepackage{graphicx}%
\usepackage{multirow}%
\usepackage{amsmath,amssymb,amsfonts}%
\usepackage{amsthm}%
\usepackage{mathrsfs}%
\usepackage[title]{appendix}%
\usepackage{xcolor}%
\usepackage{textcomp}%
\usepackage{manyfoot}%
\usepackage{booktabs}%
\usepackage{algorithm}%
\usepackage{algorithmicx}%
\usepackage{algpseudocode}%
\usepackage{listings}%

\theoremstyle{thmstyleone}%

\theoremstyle{thmstyletwo}%

\theoremstyle{thmstylethree}%

\begin{document}

\title[Article Title]{From Phytochemicals to Recipes: Health Indications and Culinary Uses of Herbs and Spices}

\author[1,2]{\fnm{Rishemjit} \sur{Kaur}}\email{rishemjit.kaur@csio.res.in}
\equalcont{These authors contributed equally to this work.}

\author[3]{\fnm{Shuchen} \sur{Zhang}}\email{szhan114@illinois.edu}
\equalcont{These authors contributed equally to this work.}

\author[1]{\fnm{Bhavika} \sur{Berwal}}\email{bhavikaberwal131@gmail.com}

\author[1,2]{\fnm{Sonalika} \sur{Ray}}\email{sonalika.csio23a@acsir.res.in}
\author[1,2]{\fnm{Ritesh} \sur{Kumar}}\email{riteshkr@csio.res.in}
\author*[3]{\fnm{Lav R.} \sur{Varshney}}\email{varshney@illinois.edu }

\affil[1]{ \orgname{CSIR-Central Scientific Instruments Organisation}, \orgaddress{\street{Sector-30C}, \city{Chandigarh}, \postcode{160030}, \country{India}}}
\affil[2]{ \orgname{Academy of Scientific and Innovative Research (AcSIR)}, \orgaddress{\city{Ghaziabad}, \postcode{201002}, \state{U.P.}, \country{India}}}
\affil*[3]{ \orgname{University of Illinois Urbana-Champaign}, \orgaddress{\city{Urbana}, \postcode{61801}, \state{IL}, \country{USA}}}


\abstract{Herbs and spices each contain about 3000 phytochemicals on average and there is much traditional knowledge on their health benefits. However, there is a lack of systematic study to understand the relationship among herbs and spices, their phytochemical constituents, their potential health benefits, and their usage in regional cuisines. Here we use a network-based approach to elucidate established relationships and predict novel associations between the phytochemicals present in herbs and spices with health indications. Our top 100 inferred indication-phytochemical relationships rediscover 40\% known relationships and 20\% that have been inferred via gene-chemical interactions with high confidence. The remaining 40\% are hypotheses generated in a principled way for further experimental investigations. We also develop an algorithm to find the minimum set of spices needed to cover a target group of health conditions. Drawing on spice usage patterns in several regional Indian cuisines, and a copy-mutate model for regional cuisine evolution, we characterize the spectrum of health conditions covered by existing regional cuisines. The spectrum of health conditions can expand through the nationalization/globalization of culinary practice.}

\maketitle

\section{Introduction}\label{sec1}

The co-evolution of plants with their pests and pathogens has led to chemical defenses in plants in the form of phytochemicals \cite{Lewis2006, PinoOtin2022, Ullah2020}. These phytochemicals, even in trace quantities, have a range of disease-alleviating properties as antioxidants, anti-inflammatories, and even anticarcinogens. Many cultures across the world have recognized the properties of these phytochemicals and have incorporated them into their culinary and medicinal practices through the use of herbs and spices, particularly in regions like China and India \cite{Laldingliani2022, Gan2023, Gonelimali, Lai2004, fraenkel1959raison, GULDIKEN201837, AhnABB2011}.

There has been some past work to understand the evolution of flavor compounds and phytochemicals in culinary practices \cite{AhnABB2011, Kinouchi_2008}. Several studies have investigated the health benefits of well-known spices and herbs, such as turmeric, saffron, fennel, and clove \cite{Mafra2021, Li2020Hong, Lopresti2014, DIAO2014109, Otunola2022, ROWLES2020158613, molecules28020887}. Although their role in disease alleviation is well known, only 63 spices and herbs are tracked by the United States Department of Agriculture (USDA) \cite{accessdata.fda.gov}. As of 2024, FooDB \cite{Harringtone026652} has listed a total of 70,926 phytochemicals and only 124 spices and herbs, out of 797 foods. 
Yet, plants have a high chemical diversity with approximately 3,000 phytochemicals or more \cite{Harringtone026652, accessdata.fda.gov}. Still, 85\% of these chemicals, which may play a role in disease prevention, remain untracked by national databases, unexplored through experimental research, and unknown to the public at large \cite{Barabasi2020}. This necessitates a systematic study of the relationships between spices-herbs, phytochemicals, and health conditions.

Researchers have developed network-based frameworks to study phytochemical-disease relationships, largely focused on a single type of phytochemical or disorder, e.g. understanding the impact of polyphenols on cardiovascular health \cite{doValle2021, Mozaffarian2021}. Tools like PhyteByte \cite{Westerman2020} and HyperFoods \cite{Veselkov2019} employ machine learning to identify cancer-fighting molecules in foods but focus solely on carcinogenic molecules. While these studies provide valuable insights, they do not comprehensively analyze the relationships between spices-herbs, phytochemicals, and health conditions. Rakhi et al. \cite{spiceRx} attempt to explore spice-phytochemical-disease relationships, but they conduct only a small-scale study with 188 ingredients yielding only 8,957 spice-disease connections. Further, they do not generate phytochemical-indication association hypotheses that can be tested experimentally. Similarly, Gao et al. \cite{Gan2023} used network approaches to study Chinese herb and disease relationships and verified their predictions with real-world patient data. However, the study is limited to Chinese herbs.

The use of ingredients varies across regional cuisines, influenced by factors such as local food availability, climatic conditions, and religious-cultural preferences \cite{RObJDefGastro}. While extensive research has been conducted on the health benefits of certain cuisines, such as the Mediterranean diet, these studies have primarily focused on macronutrients and overall dietary composition \cite{SOFI20101189}. However, the role of phytochemicals, particularly those found in herbs and spices, remains understudied in the context of regional cuisines \cite{Mediterranean}. These bioactive compounds represent a crucial aspect of the ``food as medicine" principle and have potential health benefits beyond basic nutrition. For instance, the Mediterranean diet---rich in olive oil, nuts, and various herbs---contains diverse polyphenols and other phytochemicals that have various health effects \cite{Mediterranean}.

\begin{figure}[h]
\centering
   \centering
   \includegraphics[width=\textwidth]{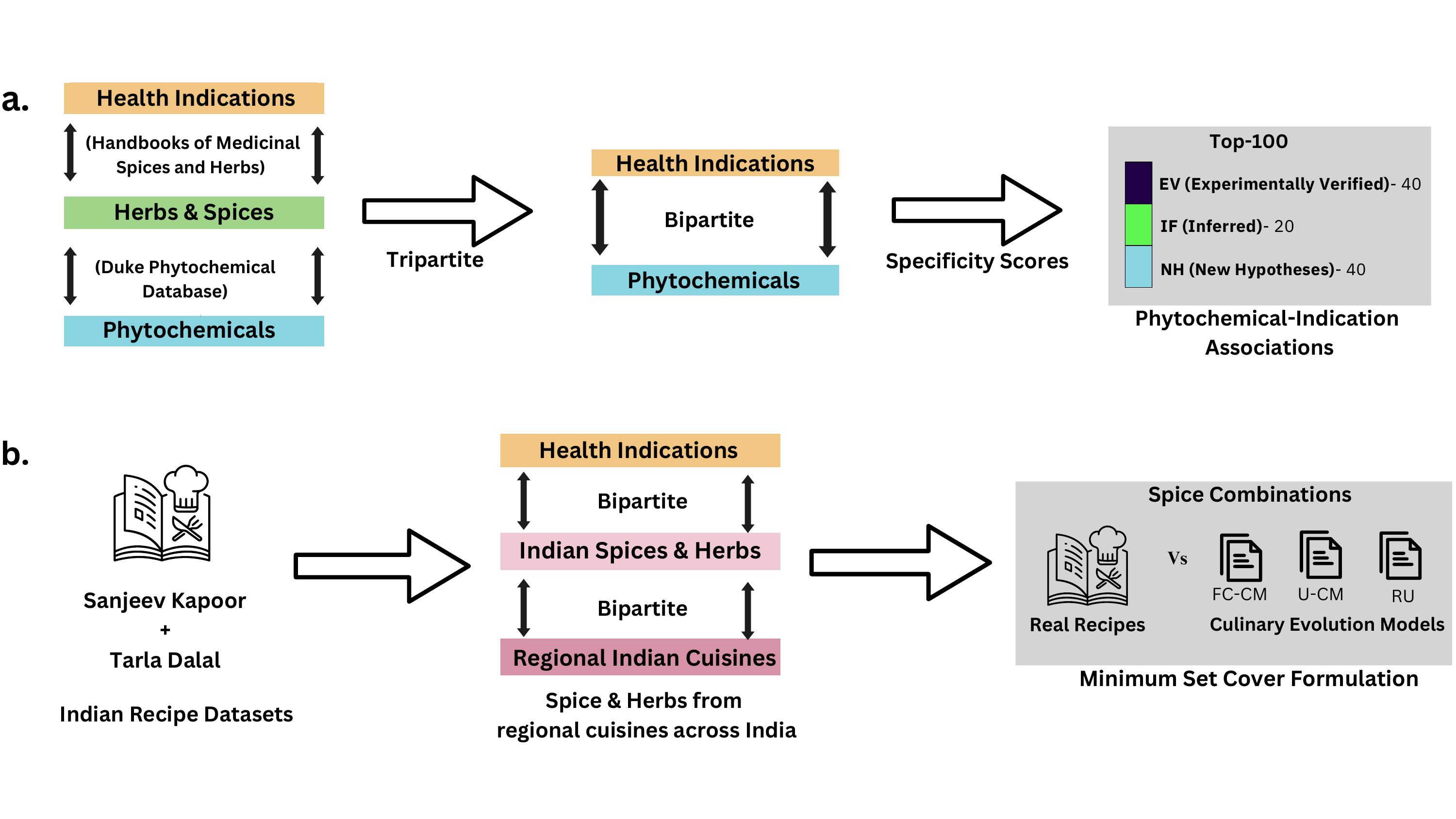} 
   \caption{Workflows representing (a) the methods used to obtain phytochemical-indication associations (b) to understand the impact of culinary globalization on health indication coverage. }
   \label{fig:workflow}
\end{figure}

The objectives of our work are twofold. First, our study aims to explore the relationships between spices and herbs, phytochemicals, and health indications using a network-based approach (Fig. \ref{fig:workflow}) and introduce a specificity score to quantify the strength of phytochemical-disease associations. We analyze 1,094 herbs and spices to generate 34,113 spice-disease relationships. We find that spices and herbs primarily address a range of health issues, including general symptoms (such as flu and fever), as well as respiratory, gastrointestinal, infectious, and musculoskeletal diseases, along with cancer. Among the top 100 inferred phytochemical-indication relationships, we rediscover 40 known associations and identify 20 that have been inferred with high confidence through gene-chemical interactions. The remaining 40 represent novel hypotheses, generated systematically, that require further experimental validation.

With globalization, there has been a notable fusion of cuisines, leading to the adaptation of traditional recipes with new ingredients. The second objective of our work is to investigate whether this culinary evolution offers any health benefits (Fig. \ref{fig:workflow}b). To address this question, we use a dataset of Indian recipes from 18 regions, as spices and herbs are integral to India's regional cuisines \cite{JainRB2015, JainB2018}. We estimate the spice and herb usage in different regional cuisines of India and then analyze the spectrum of diseases covered by different regional cuisines based on their spice usage. Further, we generate new spice combinations to simulate culinary globalization using copy-mutate models and compare their effectiveness to traditional Indian recipe spice combinations in terms of broadening health indications. Analysis of Indian cuisines showed regional trends in spice usage and highlighted that some cuisines, such as Hyderabadi, Goan, Punjabi, and Mughlai, provide better coverage of health indications. We show that randomly generated spice combinations required fewer ingredients to cover various health indications compared to traditional recipes, suggesting health benefits from culinary fusion.

\section{Results}\label{sec2}

In this section, we first present results for spice-herb, phytochemical, and indications bipartite networks, systematically analyzing the associations between health indications and phytochemicals. Second, we present results on spice usage in different Indian regional cuisines using recipe corpora, analyzing the relationships between cuisine and health indications, and comparing the minimum number of spices needed to cover a spectrum of indications using copy-mutate models.

\subsection{Network Analysis of Spices \& Herbs, Phytochemicals, and Health Indications}\label{Network2}

Two bipartite networks were created: one between spices and phytochemicals, and another between spices and health indications. From these two networks, a third bipartite network was constructed, linking indications and phytochemicals (see Fig. \ref{fig:workflow}a). The first three sub-sections provide a descriptive analysis of the three bipartite networks and the fourth section makes specific predictions of the indication-phytochemical associations.

\subsubsection{Spice-Indication Bipartite Network}\label{spiceindic2}

\begin{figure}[h]
\centering
   \centering
   \includegraphics[width=\textwidth]{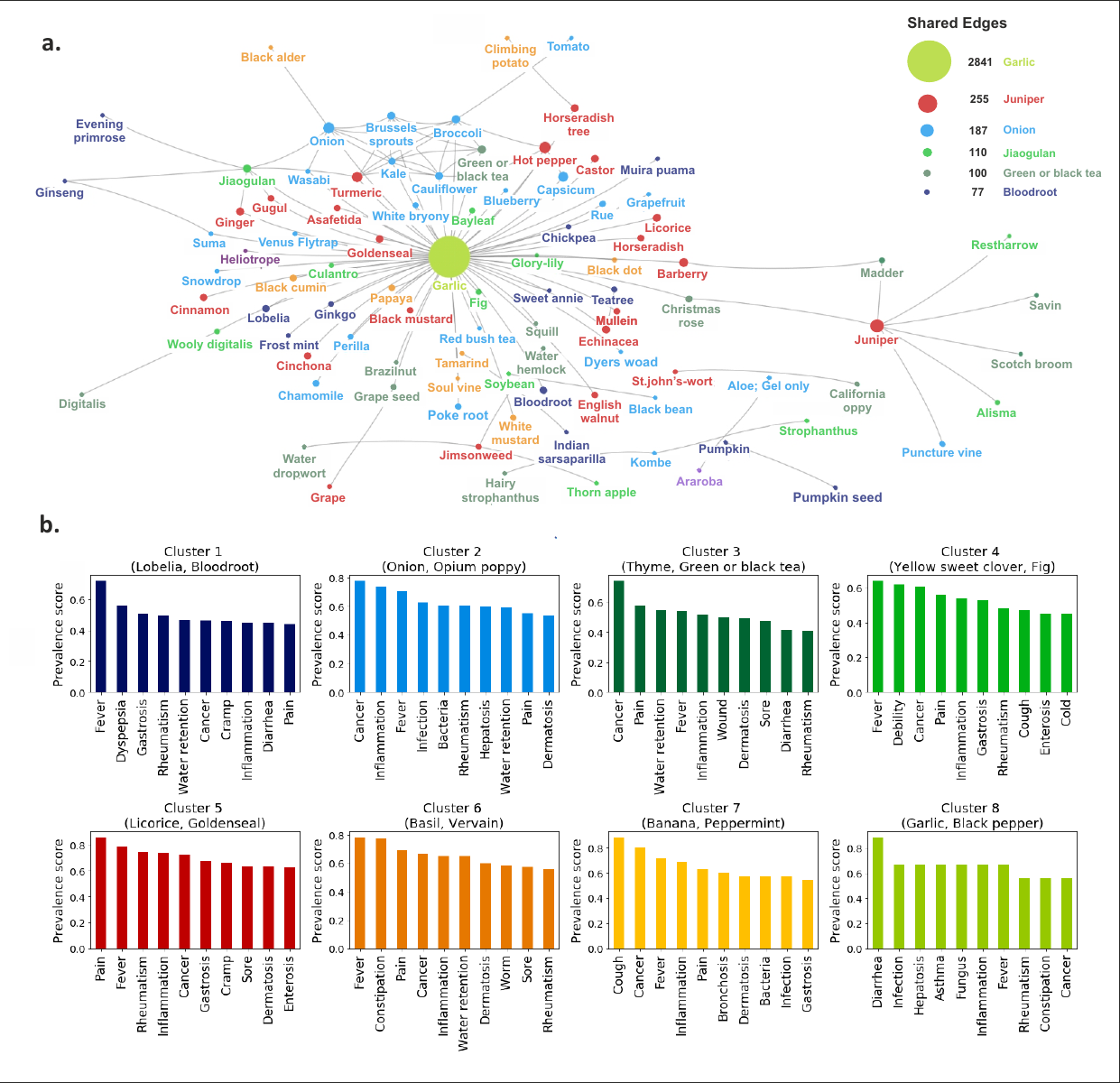} 
   \caption{(a) Backbone network visualizing connections between various spices and herbs. Each edge represents an association based on a common indication between herbs and spices. Two spices and herbs are connected if they share a common indication. The node color represents the cluster obtained from the Wakita-Tsurumi algorithm. (b) The bar plots display the prevalent indications associated with each cluster of spices and herbs.}
   \label{fig:Spice_Indication_unipartite}
\end{figure}

A bipartite network on two sets of nodes: (i) 1,094 spices and herbs and (ii) 1,597 medical indications was first built. Datasets were obtained from the Handbook of Medicinal Herbs \cite{Duke2002herbs} and the Handbook of Medicinal Spices \cite{Duke2002Spices}, providing extensive information about herbs and spices and their associated medical indications. Next, we created bipartite network projections where two nodes representing spices and herbs are connected if they share at least one indication. The Wakita-Tsurumi algorithm \cite{wakita2007finding} was applied to this bipartite projection to detect clusters of spices and herbs. For ease of visualization, we used a backbone extraction method \cite{Ghalmane2020, Yassin2023} to identify statistically significant edges as shown in Fig.~\ref{fig:Spice_Indication_unipartite}a. The central role of garlic is quite evident. To delineate the clusters, we extract bar plots of the prevalence scores of indications (see Methods~\ref{subsubsec2}) in each cluster (see Fig. \ref{fig:Spice_Indication_unipartite}b). Notice that indications belonging to different disease categories, such as respiratory ailments, gastrointestinal disorders, infectious diseases, musculoskeletal conditions, and various forms of cancer, have the highest prevalence scores across all clusters. 

Cancer is found to be the most prevalent indication in cluster 2 represented by onion and opium poppy as well as cluster 3 represented by thyme and green or black tea, suggesting they contain phytochemicals that are beneficial in cancer prevention and management. Respiratory diseases including asthma, mucososis, cough, and bronchitis are the most prevalent in cluster 7 represented by banana and peppermint, and cluster 8 represented by garlic and black pepper. Most of the clusters have a high association with at least one gastrointestinal disease.
Roughly 80\% of the spices in cluster 6, including basil and vervain, are linked to alleviating constipation. Cluster 4 has a strong association with gastrosis and hepatosis whereas cluster 8 is strongly associated with hepatosis and constipation. On the level of individual indications, pain, cough, and diarrhea are covered respectively by almost all spices within cluster 5 (represented by licorice and golden seal), cluster 7 (represented by banana and peppermint), and cluster 8 (represented by garlic and black pepper). This structured approach allows us to identify not just individual spices with therapeutic potential but also groups of spices that collectively have a range of health indications.

\subsubsection{Spice-Phytochemical Bipartite Network}\label{subsubsec2}

\begin{figure}[htbp]
\centering
   \centering
   \includegraphics[width=\textwidth]{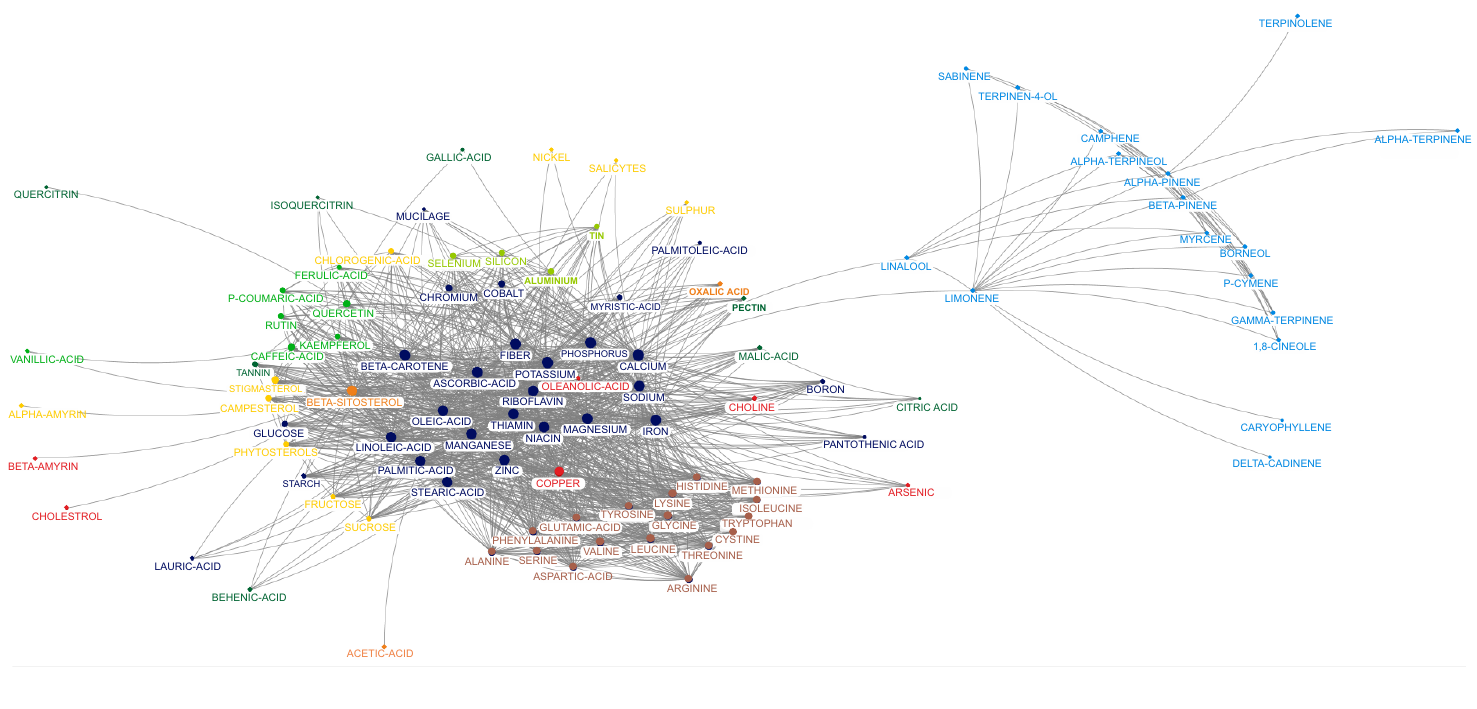} 
   \caption{Backbone network visualizing connections between various phytochemicals. Each edge represents an association based on common spices and herbs between these phytochemicals. Two phytochemicals are connected if they share a common spice/herb. The node color represents the cluster obtained from the Wakita-Tsurumi algorithm.}
   \label{fig:Phyto_Spice_unipartite}
\end{figure}

A second bipartite network between (i) 742 spices and herbs and (ii) 2,993 bioactive phytochemicals were obtained from the Duke Phytochemical Database \cite{duke1994dr} to explore relationships between spices and their constituent phytochemicals.
Fig.~\ref{fig:Phyto_Spice_unipartite} shows the projection graph on phytochemicals clustered using the Wakita-Tsurumi algorithm. The projection graph is provided in Supplementary Material (Section \ref{supplementary}).

The blue cluster on the right primarily consists of terpenes, components of essential oils derived from plants that possess anti-bacterial and anti-inflammatory properties. The light green cluster contains mostly anti-oxidants, including vanillic acid, quercetin, p-coumaric acid, and caffeic acid. The yellow cluster comprises phytosterols such as campesterol, stigmasterol, and campesterol, which are plant sterols beneficial for cardiovascular health. The Food and Drug Administration (FDA) has approved that foods containing at least 0.65 g of plant sterol esters per serving, consumed twice daily with meals for a total daily intake of at least 1.3 g, may reduce the risk of heart disease \cite{accessdata.fda.gov}.

The bottom brown cluster consists of major essential amino acids phenylalanine, methionine, leucine, histidine, and lysine, the absence of these in the diet can lead to decreased immunity, muscle loss, and even mental dysfunction.
The dark blue cluster contains a small group of polyunsaturated fatty acids (PUFAs), including linoleic acid, palmitic acid, stearic acid, and oleic acid, commonly found in oils. PUFAs boost immunity in low amounts but, consuming high amounts of PUFAs with starch can lead to diseases, particularly heart disease and weight gain.

\subsubsection{Indication-Phytochemical Bipartite Network}\label{subsubsec2}

\begin{figure}[!htbp]
\centering
   \centering
   \includegraphics[width=\textwidth]{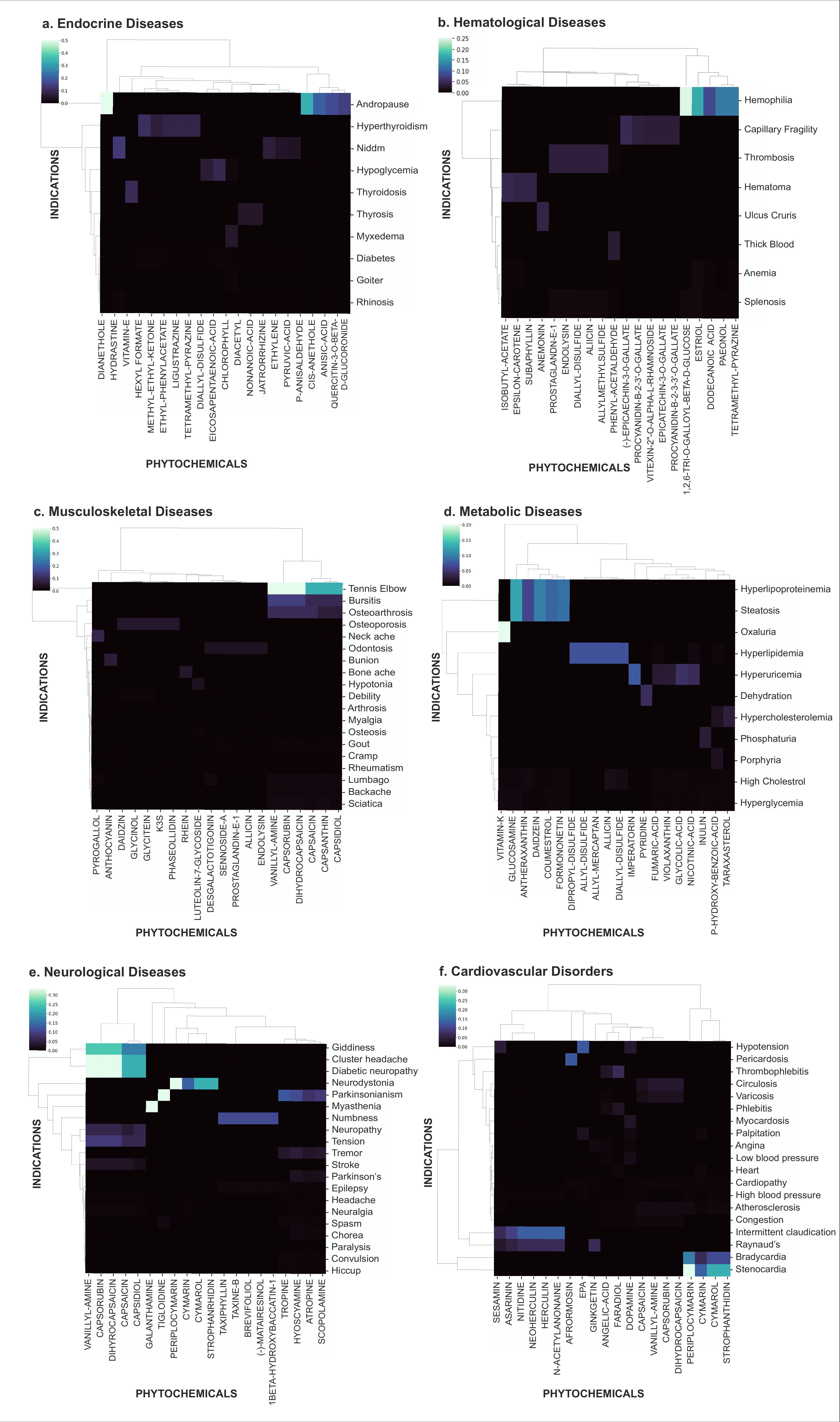} 
   \caption{Heatmaps showing the specificity scores obtained for phytochemical-indication pairs for six disease categories (a) endocrine diseases (b) hematological diseases (c) musculoskeletal (d) metabolic diseases (e) neurological diseases (f) cardiovascular diseases}
   \label{fig:Indic_Phyto_Heatmap}
\end{figure}

To understand the therapeutic properties of spices, we aimed to identify the constituent phytochemicals that contribute to their disease associations. We defined a specificity score to quantify the uniqueness of phytochemicals and their associations with indications (see Methods~\ref{subsubsubsec2}). Notice in Fig.~\ref{fig:Indic_Phyto_Heatmap}a that in endocrine diseases, high specificity was observed for dianethole and p-anisaldehyde with andropause. These phytochemicals are present in fennel and anise and are effective against endocrine diseases and other types of diseases \cite{badgujar2014foeniculum}. The efficacy of 1,2,6-tri-o-galloyl-beta-d-glucose---found in \textit{Cornus officinalis}---against protein glycation has been demonstrated, making it effective for reducing blood pressure \cite{lee2011galloyl}. Other molecules in the blood disease category did not show high specificity values.
Capsaicin and its precursor, vanillylamine, are useful as analgesics and are used in ointments for musculoskeletal pain management, which is evident in the musculoskeletal diseases specificity plots \cite{fattori2016capsaicin}. Other compounds with high specificity in this category include capsorubin and capsanthin, carotenoids found in red bell peppers that are used for pain management \cite{matsufuji1998antioxidant}, as well as dihydrocapsaicin, a compound from the same capsaicin family.
In the metabolic disease specificity plots in Fig.~\ref{fig:Indic_Phyto_Heatmap}d, high values were observed for vitamin K, glucosamine (found in many plants, including aloe vera and \textit{Cannabis sativa}), daidzein (found in soybean), coumestrol (found in soybean, spinach, brussels sprouts, and legumes), and imperatorin (found in \textit{Ammi majus} and \textit{Angelica archangelica}). Researchers have found that these molecules are effective against fatty liver, steatosis, and hyperuricemia \cite{ferramosca2017antioxidant, nasser2019effects, wroblewski2000phytoestrogens}.

Galanthamine, found in \textit{Galanthus nivalis} and other sources, showed high specificity for myasthenia gravis and Alzheimer's disease in the neurological diseases category \cite{pohanka2014inhibitors}. Similarly, high specificity scores were obtained for tigloidine, periplocymarin, cymarin, cymarol, strophanthidin, and tropine for neurological diseases like Parkinson's and neurodystonia \cite{sanghvi1968pharmacology,trautner1951tigloidine}. The high specificity scores for vanillylamine, capsorubin, and other capsaicin family molecules for cluster headaches and diabetic neuropathy are noteworthy and can be observed in Fig.~\ref{fig:Indic_Phyto_Heatmap}e. In the cardiovascular disease category, most molecules were non-specific, with some high-scoring specific molecules such as asarinin found in sesame, nitidine found in \textit{Zanthoxylum americanum}, and periplocymarin found in \textit{Strophanthus hispidus} \cite{ lee2013inhibitory, martey2014periplocymarin}. Some molecules, such as trans-isoasarone found in \textit{Acorus calamus}, show antifungal properties but are toxic and difficult to use for therapeutic purposes \cite{perrett1995anthelmintic}.

\subsubsection{Systematic Analysis of Indication-Phytochemical Associations}\label{SpecificityScore2}

\begin{figure}[h]
\centering
   \centering
   \includegraphics[width=\textwidth]{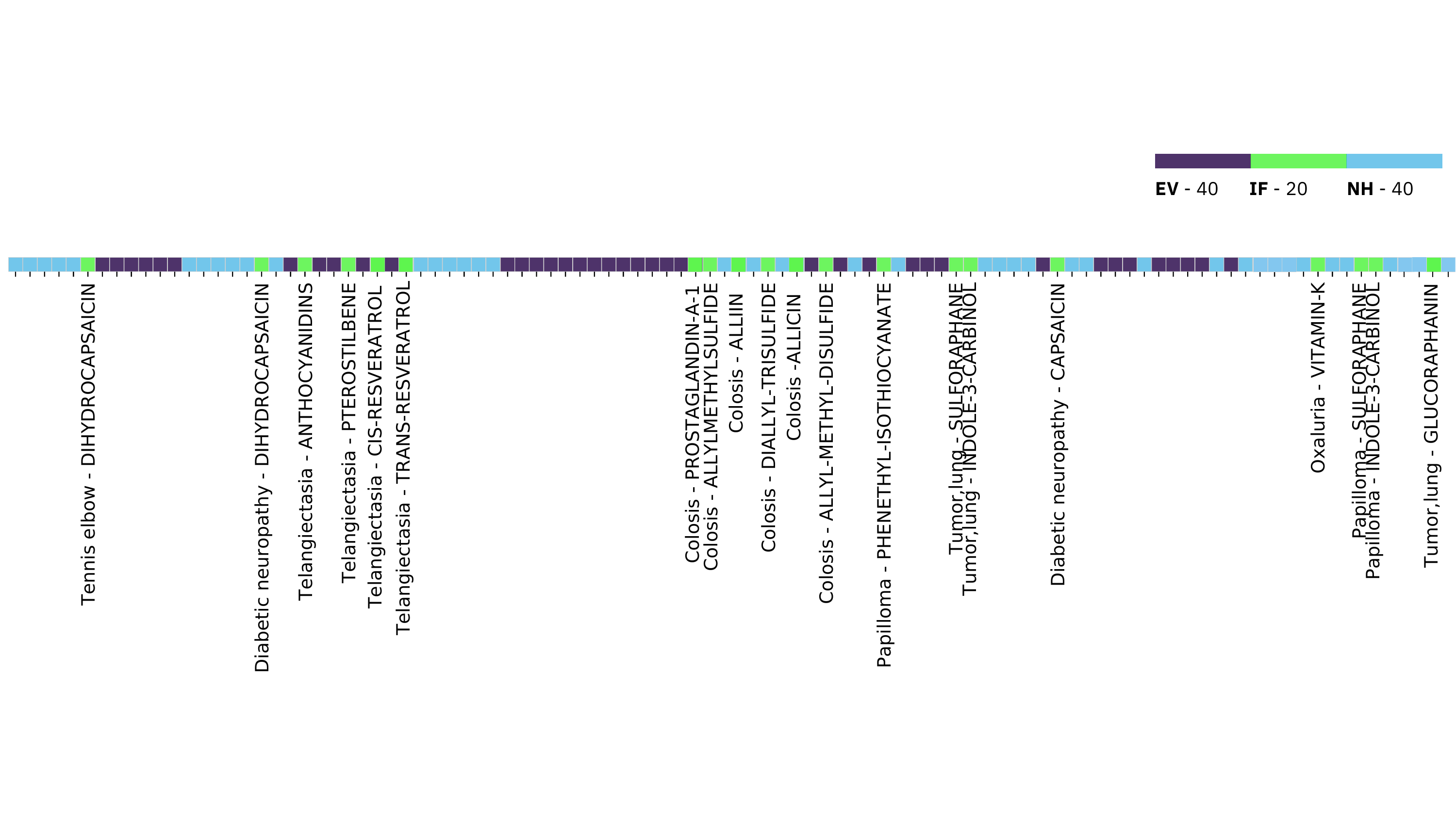} 
   \caption{The string map represents the number and nature of relationships extracted from the top 100 indication-phytochemical relationships, categorized as Inferred (IF), Experimentally Verified (EV), and New Hypotheses (NH), and sorted based on specificity scores.}
   \label{fig:CTD}
\end{figure}

To further assess the capability of the specificity score in discovering new associations and validating known relationships, we conduct a systematic analysis. We focus on the top 100 inferred indication-phytochemical relationships in terms of specificity scores.
For each inference, we first compared the results against the indication-chemical relationships provided by the Comparative Toxicogenomics Database (CTD) \cite{mattingly2003comparative}, a reliable public database containing both curated and inferred relationships. If our inferred relationships were not found in CTD, we manually searched for supporting evidence in other literature using Google Scholar. 
Out of the 100 top inferences (see Fig.~\ref{fig:CTD}), we can validate 60 indication-phytochemical relations through CTD or literature. Among these 60 inferences, 20 could be inferred through gene-chemical interactions and gene-disease associations according to CTD but have not been experimentally proven yet. The remaining 40 inferences were confirmed through experimental literature. Thus, 20 of our top inferences are new discoveries with high confidence that have also been predicted in CTD through alternative means, 40 are correct predictions backed by experimental scientific evidence, and the remaining 40 are new hypotheses that can be tested with molecular experiments. Indeed, a key use of our specificity score method is to distill novel scientific hypotheses from traditional knowledge of herbs and spices.

\subsection{Understanding Spice Usage and their Health Implications}\label{subsec2}

We use public Indian recipe corpora obtained from Sanjeev Kapoor \cite{sanjeevkapoor} and Tarla Dalal \cite{tarladalal} websites, comprising 18 regional cuisines to understand spice usage patterns across India and their association with disease categories.

\subsubsection{Similarity of Regional Cuisines of India}\label{subsubsec2}
To understand the similarity between different regional cuisines of India, we calculated the usage frequencies of different spices in each cuisine (see Methods~\ref{subsubsec2_sim}). The principal component analysis (PCA) bi-plot (Fig.~\ref{fig:Cuisines}a) of Indian cuisines and spices usage frequencies, with spices as factors projected on the principal components (PCs), reveals a clear North to South geographical orientation. The plot highlights the significant role of coconut and curry leaves in South Indian cuisines.
The cluster map (Fig.~\ref{fig:Cuisines}c) provides insights into the evolutionary relationships among Indian cuisines based on spice usage. The dendrogram shows the splitting of South Indian cuisines into distinct regional cuisines, with Andhra and Kerala cuisines exhibiting higher similarity. Gujarati and Jain cuisines, characterized by extensive use of asafoetida and the absence of onion and garlic, cluster together and share similarities with Maharashtrian and South Indian cuisines in terms of spice usage.
Punjabi and Sindhi cuisines demonstrate a lineage to Kashmiri and Mughlai cuisines, as evident from their spice usage patterns. The similarity between Kashmiri and Mughlai cuisines can be attributed to their shared use of saffron, cardamom, and clove (Fig.~\ref{fig:Cuisines}a). Similarly, Hyderabadi and Parsi cuisines show resemblance due to their pronounced use of garlic and onion.
These findings shed light on the geographical and evolutionary aspects of Indian cuisines based on spice usage. However, it is important to note that the similarities observed are based on spice usage data, and require further evidence to corroborate the cultural or historical aspects of these connections.

An interesting observation is that regional variation in spice usage across Indian cuisines correlates with genetic differences in taste perception. South Indian cuisines, like those from Kerala, Tamil Nadu, and Andhra Pradesh, use more bitter-tasting ingredients such as mustard seeds and curry leaves (see Fig.~\ref{fig:Cuisines}a). This culinary practice aligns with genetic data showing a higher frequency of the rs2274333 A allele in the CA6 gene in southern Indian populations, which is associated with increased sensitivity to bitter tastes. Conversely, northern Indian cuisines, including those from Punjab, Kashmir, and Parsi traditions, tend to emphasize aromatic spices such as cardamom, saffron, and complex spice blends like garam masala (see Fig.~\ref{fig:Cuisines}a). This preference for aromatic over bitter flavors corresponds to a lower frequency of the rs2274333 A allele in northern populations, suggesting a potentially reduced sensitivity to bitter tastes in these regions \cite{indiantasteprofile}. These patterns indicate a possible relationship between genetic predisposition to taste sensitivity and regional culinary practices. However, cuisine development is influenced by various factors like cultural history, ingredient availability, and environmental conditions, in addition to genetic factors \cite{indiantasteprofile, bittertaste}.
    
\subsubsection{Authentic Spices for Every Cuisine}\label{subsubsec2}

\begin{figure}[h]
\centering
   \centering
   \includegraphics[width=\textwidth]{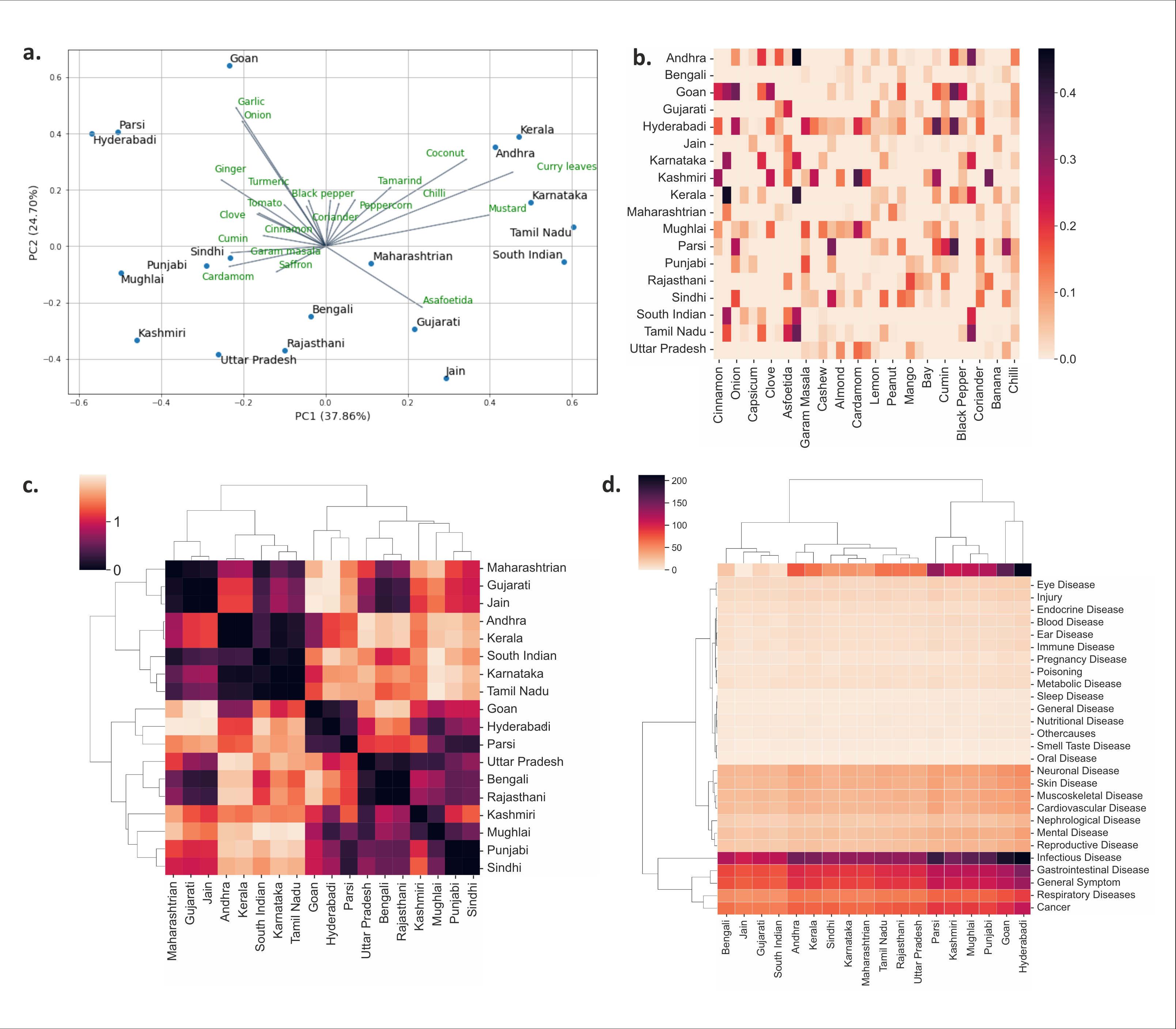} 
   \caption{(a) PCA bi-plot obtained from the frequency of spice usage in different cuisines representing regions as scores and spices as loadings. PC1 and PC2 account for 37.86\% and 24.79\% of the total variance respectively.
   (b) Heatmap showing the authenticity of spices in Indian cuisines. The darker regions indicate more frequent use of certain spice pairs. (c) A cluster map was obtained by calculating the cosine similarities between the PCs of different regions. Cuisines that are closer together in the dendrogram have more similar spice usage profiles. (d) Cluster heatmap showing the indication coverage for the different regional cuisines of India.}
   \label{fig:Cuisines}
\end{figure}

Fig.~\ref{fig:Cuisines}b presents a heat map of the authentic spices, defined by their unique use in each regional Indian cuisine. While there is a substantial overlap in spice usage across cuisines, the analysis reveals several interesting observations, some well-known and others less so.
The presence of asafoetida in Jain and Gujarati cuisines is well-established, as Jains and many Gujaratis exclude onion and garlic from their diet for religious reasons, but asafoetida contains di-allyl sulfur, the same pungent phytochemical as in garlic and onion, making it an ideal substitute \cite{Batiha2020}.
As noted earlier, curry leaves and coconut are integral to South Indian cuisines. Mughlai and Hyderabadi cuisines also heavily feature cardamom and clove, while Kashmiri cuisine is uniquely characterized by the presence of saffron and fennel.
A lesser-known fact is the use of peanuts as an authentic spice/herb in Maharashtrian cuisine, which is not widely recognized. These findings highlight the diversity and complexity of Indian cuisines, showcasing the interplay of regional preferences, religious influences, and unique spices that define the authentic flavors of each culinary tradition.

\subsubsection{Relationship between Cuisine and Indications}\label{subsubsec2}
 
Fig.~\ref{fig:Cuisines}d presents a heatmap with hierarchical clustering of Indian cuisines based on their indication coverage. It shows that regional cuisines have better coverage for five disease categories: cancer, respiratory diseases, general symptoms, gastrointestinal diseases, and infectious diseases. Hyderabadi, Goan, Parsi, Punjabi, and Mughlai cuisines show a broader and stronger coverage of the indication spectrum than the other cuisines.  The analysis reveals that Hyderabadi and Goan cuisines exhibit the highest scores for alleviating infectious diseases, followed closely by Parsi cuisine.

\subsubsection{Min Set Cover Formulation}\label{subsubsec2}

Each cuisine has a unique profile of herbs and spices with some having combinations with greater disease mitigation, as observed in Mughlai and Hyderabadi cuisine. New fusion cuisines have emerged with increasing globalization as ingredients from different cultures are blended to create new recipes. Here, we study how well combinations of spices from culinary practice cover a spectrum of diseases, using a minimum set cover algorithm to find the minimum set of spices required to cover a range of indications for each disease category. 

We then compare the disease coverage capability of recipes generated under four different settings: real settings (using recipes from Tarla Dalal and Sanjeev Kapoor) and three random settings. The random settings simulating culinary globalization include the uniform copy mutate (U-CM) model, the frequency conserved copy mutate (FC-CM) model \cite{JainB2018, Kinouchi_2008}, and the random uniform (RU) model (see Methods~\ref{Randonrecipe2}). To ensure a fair comparison against the real recipes, each recipe in the random settings contained six spice ingredients, which corresponds to the median number of spices per recipe in the real recipe dataset. We generated 50 sets of 5,636 recipes for each random model and used the mean size of the minimum recipe sets for comparison.

\begin{figure}[h]
\centering
   \centering
   \includegraphics[width=\textwidth]{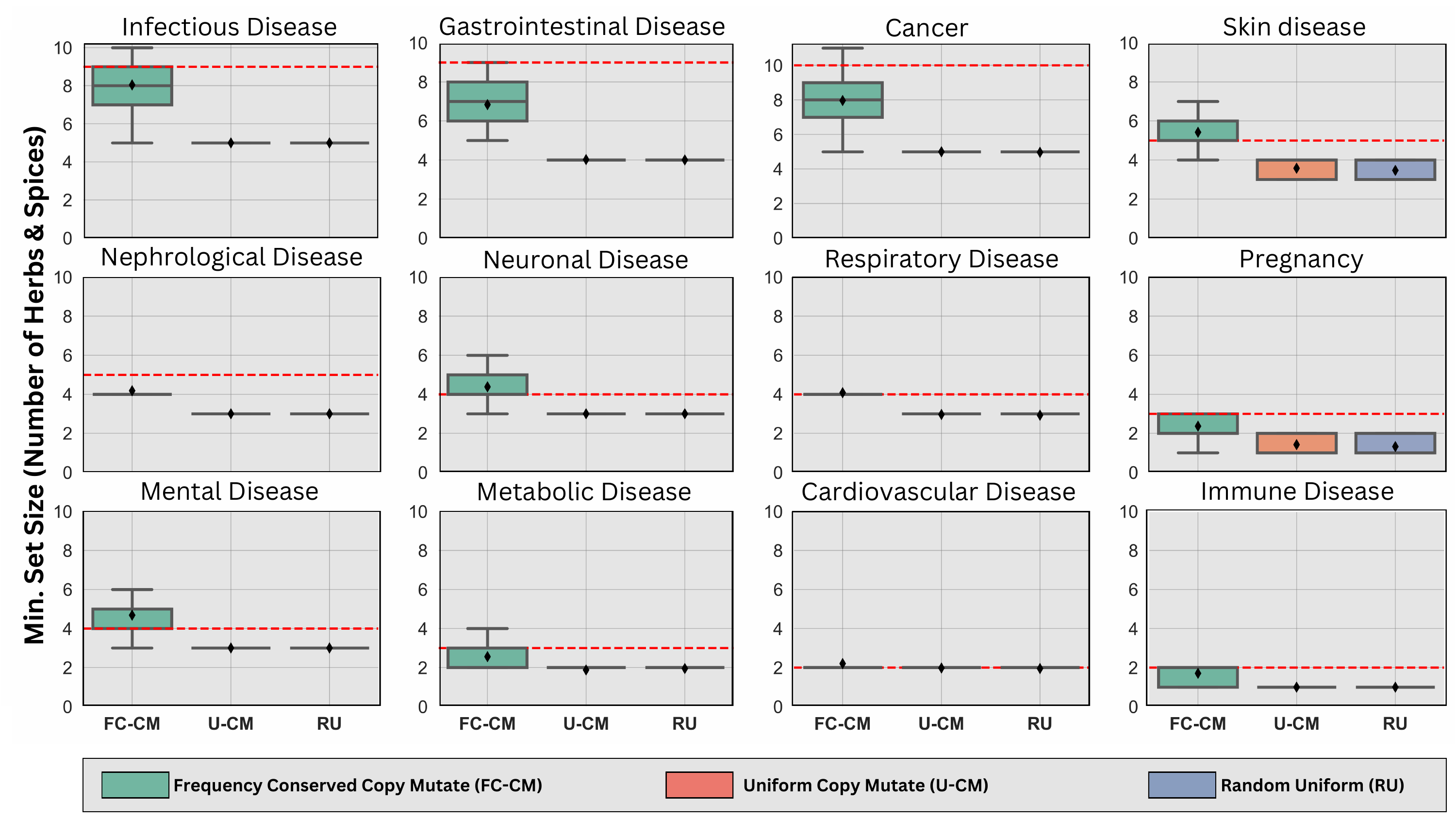} 
   \caption{Each plot represents the distribution of the minimum number of spices obtained using the minimum set cover algorithm for the three copy-mutate models (FC-CM, U-CM, RU) for 12 disease categories. The red line represents the size of the minimum set of spices obtained for the recipe dataset.}
   \label{fig:minset_boxplot_disease}
\end{figure}

Fig.~\ref{fig:minset_boxplot_disease} compares the mean size of the minimum set of spices needed to cover health indications in both the random and the actual settings (obtained from the recipe datasets) for 12 different disease categories. Note that the size of the minimum set of spices under FC-CM is close to that of the original recipe datasets, which can be attributed to the fact that it conserves the frequency of spices used. For most disease categories, fewer spices are needed to cover the spectrum of indications for U-CM and RU models. These randomly generated recipes require fewer spices to cover infectious, gastrointestinal, and cancer indications than traditional Indian cuisines. This efficiency may be due to spices like garlic that address multiple health concerns and are used extensively in various Indian regional cuisines. However, actual spice usage in cuisines is also influenced by flavor, ingredient interactions, availability, and cultural factors, not just health benefits. Going forward, it is of interest to explore mixing while preserving or enhancing the flavor of the recipes \cite{Varshney2019}.

\section{Discussion}\label{sec3}

We took a network-driven approach to discover knowledge regarding spices and herbs, their constituent phytochemicals, health indications, and their culinary usage. While spices and herbs cover many disease categories, general symptoms, respiratory disease, gastrointestinal disease, infectious disease, musculoskeletal disease, and cancer are the most widely covered. From the spice-indication and spice-phytochemical bipartite networks, a third bipartite between phytochemicals and indications was constructed.
We then probed deeper into the indication-phytochemical bipartite by defining specificity scores which indicated the degree of association between the indications and phytochemicals. 

Various indication-phytochemical associations emerged out of our analysis, such as dianethole, p-anisaldehyde found in fennel, and anise which are useful for treating endocrine diseases. For better understanding, we performed a systematic analysis of indication-phytochemical associations by ranking them based on specificity scores. The analysis of the top 100 inferences showed that we achieved a high confidence score within 60 associations and some of the inferred relationships are yet unknown. This may prove to be a new method for the identification of new phytochemical-disease associations. The known relationships were verified against common gene-chemical and gene-disease associations and thus are biologically relevant.The remaining 40 associations of the top inferences labeled `NH' are the new predicted associations that could be tested experimentally.
Compared to SpiceRx \cite{spiceRx} our study provides deeper insights by analyzing 1,094 herbs and spices to generate 34,113 spice-disease relationships. While the past work does use existing databases like PhenolExplorer \cite{phenolexplorer} and CTD \cite{mattingly2003comparative} to find phytochemical-diseases, we introduced a normalized specificity score to quantify the strength of phytochemical-disease associations, providing higher granularity in assessing unique spice-phytochemical-diseases relationships. Our methods thus, address these limitations, providing a more comprehensive and quantitative analysis of spice-disease interactions.

The analysis of Indian cuisines provides insights into the spice usage and geography of the country. There is a trend in the spice usage from North to South. We also found authentic spices in the cuisines and found that spices like asafoetida, curry leaves, cardamom, and clove are important ingredients in the Jain, South Indian, Hyderabadi, and Mughlai cuisines. We further looked into the spice usage in Indian cuisines and their disease associations. One important observation coming out of this analysis is that the spices/herbs used do not cover all disease categories and it would be beneficial to add other kinds of spices/herbs to the diet. This analysis also indicated that some cuisines like Hyderabadi, Goan, Punjabi, and Mughlai provide better coverage of indications than the other cuisines. It points out that mixing some cuisines could further help in indication coverage, as the minimum number of spices needed to cover the different kinds of diseases is quite small for randomly generated recipes as compared to the real recipes. This highlights that the fusion of cuisines may allow for a diverse diet and enhance the health benefits of our diet. However, the cultural significance of spices and herbs cannot be understated. Globalization may break traditions, but it might have positive impacts too.  For example, AI-generated recipes can be quite compelling.

To conclude, the contributions of this paper are twofold. Firstly, we used a network-based approach to develop a method for scientific phytochemical-health hypothesis generation by distilling traditional knowledge of herbs and spices. Then we systematically analyzed the health impacts of culinary traditions and how the global evolution of food may cover more health indications.

\section{Methods}\label{sec3}

\subsection{Data, Preprocessing, and Categorization}\label{subsubsec2}

We collected the medicinal indications of herbs and spices from Handbook of Medicinal Spices \cite{Duke2002Spices} and Handbook of Medicinal Herbs \cite{Duke2002herbs}. These handbooks provide medicinal information for a large collection of essential spices/herbs along with their cultivation and chemistry. Each spice/herb is associated with a list of indications. There are 1,094 spices/herbs and  1,597 indications associated with them. We have also collected 2,993 constituent phytochemicals of these herbs and spices from the Duke Phytochemical Database \cite{duke1994dr}. 
The indications were labeled into 24 categories using the International Classification of Diseases (ICD‐11) \cite{ICD} and Malacards \cite{RappaportNST2013}. These categories include gastrointestinal, respiratory, cancer, infectious, mental, reproductive, and cardiovascular diseases, among others. In cases where an indication was not found in either database, we searched for its synonyms using web searches and repeated the categorization process. Note that some indications may belong to multiple categories.

To investigate the role of spices in Indian cuisine, we collected recipe data from two popular culinary websites in India by Tarla Dalal and by Sanjeev Kapoor. These websites feature a diverse collection of recipes from various Indian and international cuisines. Our combined dataset comprises 13,212 recipes in total, with 3,876 recipes sourced from Tarla Dalal \cite{tarladalal} and 9,336 recipes from Sanjeev Kapoor \cite{sanjeevkapoor}. Each recipe entry includes details such as the ingredients used, their quantities, preparation methods, and the associated cuisine or geographical region. 

For our analysis, we focused specifically on recipes belonging to Indian cuisines, such as Punjabi, Bengali, and others. We filtered the dataset to include only Indian cuisine recipes containing spices and herbs, resulting in a final dataset of 5,636 recipes, with 3,595 recipes from Tarla Dalal and 2,041 recipes from Sanjeev Kapoor.

For systematic analysis to understand the phytochemical-indication association, we used the Comparative Toxicogenomics Database (CTD) \cite{mattingly2003comparative} and published literature. CTD is a reliable public database containing both curated and inferred relationships. The curated relationships are extracted from published literature by CTD curators, while inferred relationships are established through CTD-curated chemical-gene interactions. In the latter case, chemical A is ``inferred" to be associated with disease C via gene B if chemical A directly interacts with gene B, and gene B is associated with disease C. We classified the results as \textit{Inferred} if we found a chemical-gene interaction in CTD. If not, then we looked into literature. Unless we found an associated phytochemical-disease relationship, we classified it as \textit{Experimentally Verified}. If it was not found in either, we classified it as \textit{New Hypotheses}.

\subsection{Network Construction and Visualization}\label{subsubsec2}
We created three bipartite networks to understand the relationships between spices/herbs, indications, and phytochemicals: spice-indication, spice-phytochemical, and indication-phytochemical. For each bipartite network $G = (U, V, E)$ on two disjoint sets of nodes $U$ and $V$, projection graphs on both $U$ and $V$ are generated. In a projection graph on set $U$, two nodes are connected if they are both connected to at least one node in $V$. The bipartite networks can also be represented by a bi-adjacency matrix $B$, where the rows represent the nodes of $U$, and the columns represent the vertices of $V$. In this matrix, $B_{ij} = 1$ if there is an edge between vertex $i$ of $U$ and vertex $j$ of $V$, and $0$ otherwise. We can project this bi-adjacency matrix to a unipartite projection by calculating $BB^T$.

The spice-indication and spice-phytochemical bipartite networks were derived directly from the Handbooks of Medicinal Spices and Herbs \cite{Duke2002Spices, Duke2002herbs} and the Duke Phytochemical Database \cite{duke1994dr}, respectively. Indication-phytochemical associations were inferred through an integration of the spice-phytochemical and spice-indication bipartite networks. The indication-phytochemical bipartite network was constructed by linking each constituent phytochemical of a spice directly to all the indications the spice is associated with, and weak links were filtered out.

For visualization purposes, we used a backbone extraction algorithm to abstract the network while preserving the statistically significant links and nodes \cite{Ghalmane2020, Yassin2023}. The weight of each link was calculated as the total number of shared nodes, and the size of each node was proportional to the number of its connections within the backbone graph. The Wakita-Tsurumi algorithm \cite{wakita2007finding} was applied to the full unipartite network to identify groups of nodes that are close in the network topology.

\subsection{Prevalence Score}\label{subsubsec2}
The Wakita-Tsurumi algorithm is applied to the spice-indication and spice-phytochemical bipartite networks to identify spice clusters with similar therapeutic properties and chemical compositions, respectively. To characterize each cluster, we introduce the prevalence score, which quantifies the importance of each indication or phytochemical within a cluster. For a spice cluster $K_i$, the prevalence score of an indication or phytochemical is calculated by counting the number of spice links associated with it within the cluster in the respective bipartite network and then normalizing the count by dividing it by the total number of spices in the cluster ($|K_i|$). This normalization ensures fair comparisons across clusters of different sizes.

A higher prevalence score indicates a stronger association between an indication or phytochemical and the spices in the cluster, suggesting its importance within that group. Examining prevalence scores within each cluster helps identify key therapeutic properties or chemical compounds characterizing the spices.

\subsection{Specificity Score}\label{subsubsubsec2}

To quantify the uniqueness of phytochemicals in their association with a certain disease, we measured a specificity score for each indication-phytochemical pair $(u_i, v_j)$. The specificity score is computed as the number of links between them, normalized by the product of the count of $u_i$'s spice associations and the count of $v_j$'s spice associations:

\begin{equation}
S_{ij} = \frac{e_{ij}}{e_i . e_j}
\label{eq:Specificity score}
\end{equation}

\noindent{where $e_{ij}$ is the number of edges between nodes $u_i$ and $v_j$, and $e_i$ and $e_j$ are the total number of edges connected with nodes $u_i$ and $v_j$, respectively.}

\subsection{Usage and Authenticity of Spices}\label{subsubsec2_sim}

The authenticity score \cite{AhnABB2011} measures how unique or representative a spice or herb is to a specific cuisine compared to its usage in other cuisines.

The usage frequency $F_i^c$ captures the proportion of recipes within a cuisine $c$ that include the spice or herb $i$. and it is defined as:

\begin{equation}
F_i^c = \frac{n_i^c}{N^c}
\label{eq:Usage_Freq}
\end{equation}

\noindent{where $n_i^c$ represents the number of recipes in cuisine $c$ that contain spice or herb $i$ and $N^c$ represents the total number of recipes in cuisine $c$. A higher value of $F_i^c$ indicates that the spice or herb $i$ is more commonly used in cuisine $c$. Next, we calculate the average usage frequency of spice or herb $i$ across all other cuisines except $c$:}

\begin{equation}
\langle F_i^{c'} \rangle_{c' \neq c} = \frac{\sum_{c' \neq c} F_i^{c'}}{|C| - 1}
\end{equation}

\noindent{where $c'$ represents all cuisines other than $c$ and $|C|$ represents the total number of cuisines. The term $\langle F_i^{c'} \rangle_{c' \neq c}$ represents the average usage frequency of spice or herb $i$ in all cuisines other than $c$. It provides a baseline for comparing the usage of spice or herb $i$ in cuisine $c$ to its usage in other cuisines. Finally, the authenticity score of spice or herb $i$ in cuisine $c$ is defined as:}

\begin{equation}
A_i^c = F_i^c - \langle F_i^{c'} \rangle_{c' \neq c}
\end{equation}

\noindent{The authenticity score $A_i^c$ measures the relative usage frequency of spice or herb $i$ in cuisine $c$ compared to its average usage frequency in all other cuisines. A positive value of $A_i^c$ indicates that spice or herb $i$ is used more frequently in cuisine $c$ than in other cuisines, suggesting that it is more authentic or representative of cuisine $c$. Conversely, a negative value of $A_i^c$ indicates that spice or herb $i$ is used less frequently in cuisine $c$ compared to other cuisines, suggesting that it is less authentic or representative of cuisine $c$.}

\subsection{Indication Coverage in Cuisines}\label{indic_coverage}

 The spice-indication matrix $M_{SI}$, capturing the association between spices and their indications is defined as:

\begin{equation}
M_{SI}[s,i] = 
\begin{cases}
    1 & \text{if spice } s \text{ is associated with indication } i \\
    0 & \text{otherwise}
\end{cases}
\end{equation}

\noindent{The cuisine-spice matrix $M_{CS}$ represents the usage of spices in different cuisines as defined in equation \ref{eq:Usage_Freq}. The cuisine-indication matrix $M_{CI}$ representing the indication coverage of cuisines is calculated as:}

\begin{equation}
M_{CI} = M_{CS}\times M_{SI}
\end{equation}

\noindent{Each element of $M_{CI}$ represents the strength of association between cuisine $c$ and indication $i$.}

\subsection{Minimum Set-Cover Algorithm}\label{minset}

To find the minimum subset of spices that can cover a specific set of indications, we use the minimum set cover problem on a bipartite network.
This approach helps in understanding the efficiency and optimization of spice usage in regional cuisines for potential health benefits. Given a bipartite network $G = (U, V, E)$, where $U$ and $V$ are two disjoint sets of nodes representing spices and diseases, respectively, and $E$ is the set of edges connecting them, our goal is to identify the smallest subset of nodes in $U$ that covers all the nodes in a specified subset of $V$.

\begin{equation}
\text{Minimize } \mathbf{c}^T \mathbf{X}
\label{eq:minsetcover}
\end{equation}
\begin{align}
\text{Subject to:} \quad & 0 \leq \mathbf{X}_j \leq 1 \quad \text{for } j = 1, 2, ..., |V| \nonumber \\
& \sum_j \mathbf{B}_{ij}\mathbf{X}_j \geq 1 \quad \text{for } i = 1, 2, ..., |U| \nonumber
\end{align}

\noindent{We approach this problem by formulating it as an integer linear programming problem. Let $|U|$ and $|V|$ denote the number of nodes in sets $U$ and $V$, respectively. The problem was formulated with the constraints expressed as:}

\begin{equation}
\mathbf{G}\mathbf{X} \leq \mathbf{h}
\label{eq:constraint}
\end{equation}
where,
\begin{align}
\mathbf{G} &= [\mathbf{B}, \mathbf{I}_{|V|}, -\mathbf{I}_{|V|}]^T \\
\mathbf{h} &= [-\mathbf{1}_{|U|}, \mathbf{1}_{|V|}, \mathbf{0}_{|V|}]^T
\end{align}

 $\mathbf{B}$ is the adjacency matrix of the bipartite network with dimension $|V| \times |U|$, where $|U|$ and $|V|$ are the total number of nodes in sets $U$ and $V$, respectively. Each entry of $\mathbf{B}$ is either 1 or 0, indicating whether or not there exists at least one link between each node pair between $U$ and $V$. Here, $\mathbf{X}$ is a vector of size $|V|$ to be solved, and each entry of $\mathbf{X}$ is either 1 or 0, indicating whether or not the node from set $U$ should be included in the minimum set cover. 
We obtain a fractional solution for X, which is then rounded to obtain a feasible solution to the original minimum set cover problem. The rounded solution represents the minimum subset of spices (nodes from $U$) that covers all the diseases (nodes in the specified subset of $V$) in the bipartite network $G$.

\subsection{Random Recipe Generation}\label{Randonrecipe2}
We constructed three random recipe datasets using different models: frequency-conserved copy-mutate (FC-CM), uniform copy-mutate (U-CM), and random uniform (RU) \cite{JainB2018, Kinouchi_2008}. These models aim to simulate the process of recipe creation and evolution over time. The FC-CM algorithm begins by creating an initial random pool $I_0$ of 10 spice ingredients and a seed pool $R_0$ of 20 recipes. Each recipe in $R_0$ is generated by randomly selecting $S = 6$ spice ingredients from $I_0$. The value of $S$ is chosen based on the median number of spice ingredients per recipe in the real recipe dataset. Each spice ingredient in $I_0$ is assigned a fitness value based on its empirical frequency in the real data, reflecting how frequently it is used across different cuisines. Spice ingredients with higher frequencies in real-world recipes are considered more fit and versatile, and thus have a higher chance of being selected for a recipe.

At each time step, a mother recipe is randomly selected from the recipe pool $R_0$. A copy of this mother recipe is made, and the copy undergoes a mutation process. During mutation, an ingredient with fitness $f_i$ is randomly chosen from the copied recipe and compared with another ingredient with fitness $f_j$, which is randomly selected from the ingredient pool $I_0$. If $f_j > f_i$, the old ingredient $i$ is replaced with the new ingredient $j$. This mutation process is repeated $M = 6$ times, after which the mutated copy recipe is added back to the recipe pool $R_0$, becoming a potential candidate for selection as a mother recipe in the next time step. To maintain a diverse ingredient pool, the ratio $r$ between the size of the ingredient pool $I_0$ and the size of the recipe pool $R_0$ is checked at the beginning of each time step. If $r$ falls below a threshold of 0.2, new spice ingredients are introduced to $I_0$ by randomly selecting from the list of all available spice ingredients. These new ingredients are added to the existing pool, expanding the variety of spices available for recipe creation. The FC-CM process continues until the desired number of recipes is reached, which in this case is 5,636, matching the number of recipes in the real recipe dataset.

The U-CM model follows a similar copy-mutate process as the FC-CM model, with the key difference being that no fitness value is assigned to each spice ingredient. In the U-CM model, each spice has an equal probability of being selected for a recipe, and the mutation process involves replacing a randomly chosen ingredient with another randomly selected ingredient from the pool.

In the RU model, recipes are constructed by randomly choosing ingredients with uniform probability, without considering any fitness values or mutation processes.

\bibliography{sn-bibliography}

\newpage
\maketitle

\section{Supplementary Material}\label{supplementary}

\subsection{Spice-Phytochemical Biparite Graph}

\begin{figure}[!h]
\centering
   \centering
   \includegraphics[width=\textwidth]{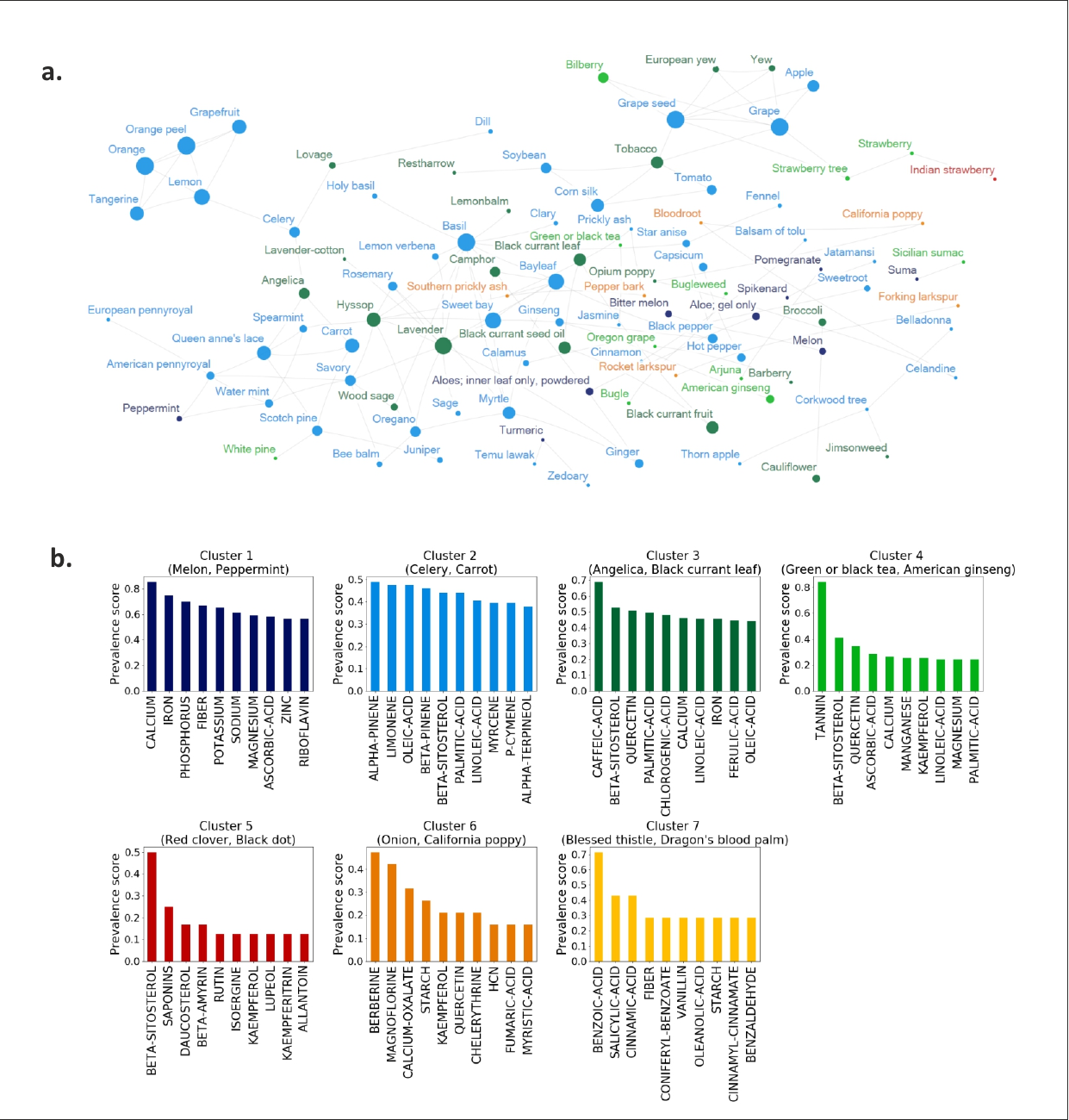} 
   \caption{(a) Backbone network visualizing connections between various spices and herbs where each edge represents an association based on a common phytochemical between herbs and spices. Two spices and herbs are connected if they share a common phytochemical. The node color represents the cluster obtained from the Wakita-Tsurumi algorithm. (b) Bar plots showing prevalent phytochemicals associated with each cluster of spices and herbs.}
   \label{fig:Spice_Phytochemical_Unipartite}
\end{figure}

The uniparite projection between spices and phytochemicals is represented (see Fig.~\ref{fig:Spice_Phytochemical_Unipartite}) with the size of the node representing the prevalence of a particular phytochemical within the spice. The connectivity between spices indicates similarities in their properties. For instance turmeric and ginger are connected, which suggests they share certain phytochemicals like curcuminoids in turmeric, known for their anti-inflammatory properties, which might also be present in ginger.

The cluster representing spices like celery, lemon, bayleaf, and juniper (in blue) is observed all across the unipartite. 
The bar plots in Fig.~\ref{fig:Spice_Phytochemical_Unipartite}b illustrate the prevalence of specific phytochemicals across seven distinct spice clusters. Cluster 1 (melon, peppermint) exhibits a high prevalence of `alpha-tocopherol' (vitamin E) and `calcium,' suggesting potential antioxidant and bone health benefits. Cluster 2 (celery, carrot) is rich in `beta-sitosterol', a phytochemical with potential cholesterol-lowering effects, indicating possible cardiovascular benefits. Cluster 3 (angelica, black currant leaf) contains `caffeic acid', known for its anti-inflammatory properties, implying potential benefits in managing inflammatory conditions. Cluster 4 (green or black tea, American ginseng) shows a high prevalence of `tannin', associated with antioxidant benefits and possibly protective effects against certain chronic diseases. Cluster 5 (red clover, black dot) features `biochanin A', which has been studied for its potential estrogenic and bone health benefits, suggesting these spices might be beneficial in hormone-related health issues. Cluster 6 (onion, California poppy) exhibits a prevalence of `quercetin', a flavonoid with well-documented antioxidant and anti-inflammatory effects, indicating potential support for overall health and well-being. Lastly, Cluster 7 (blessed thistle, Dragon's blood palm) prominently features `cinnamic acid' and its derivatives, suggesting possible use in managing blood sugar and offering antimicrobial benefits.

\subsection{Indication-Phytochemical Associations}

\begin{figure}[!h]
\makebox[\textwidth][c]{\includegraphics[width=2.0\textwidth]{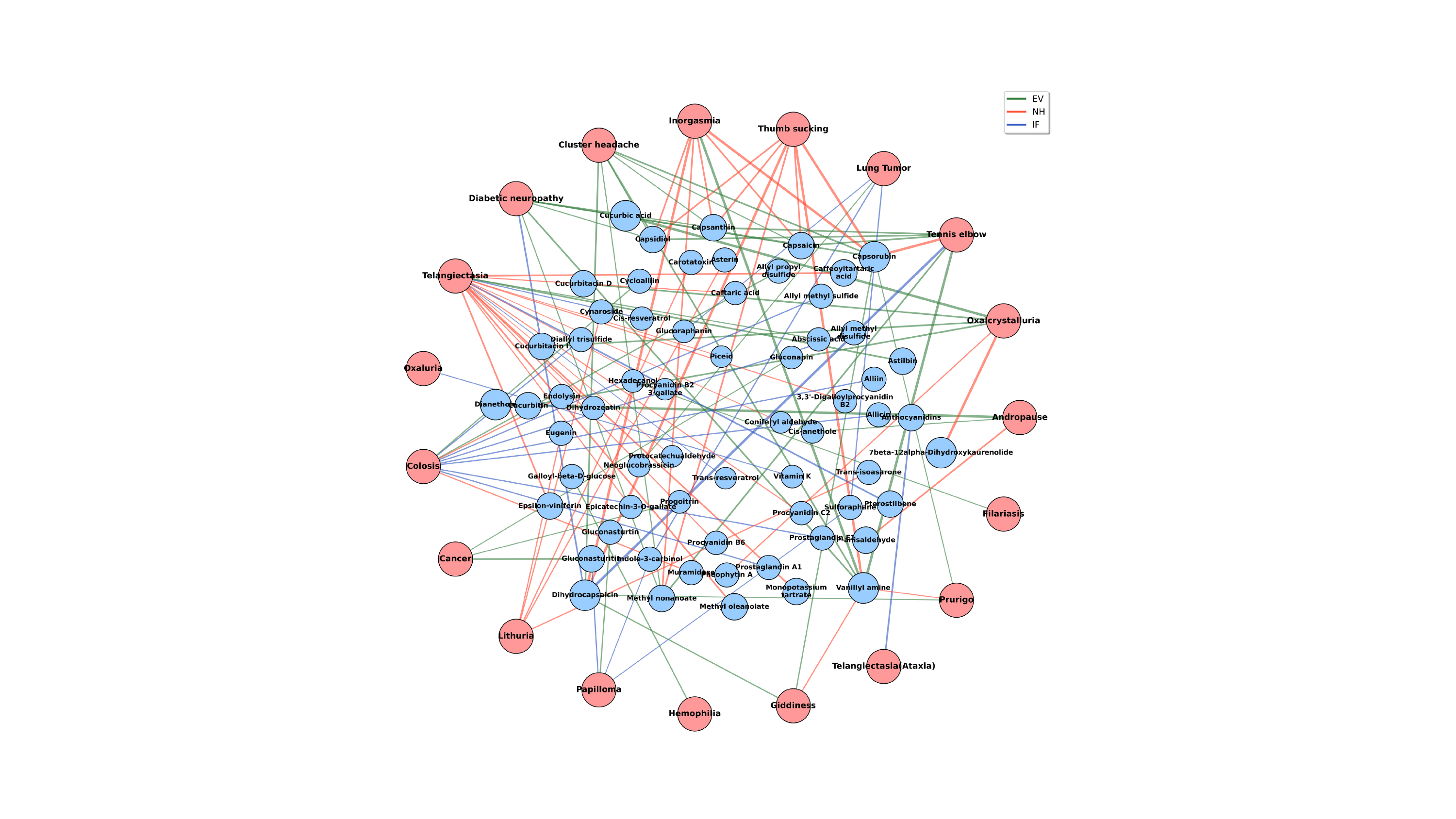}}
\caption{Bipartite network of phytochemical-indication associations. Blue nodes represent phytochemicals and red nodes represent indications. Edge colors denote association types: Inferred (IF), Experimentally Verified (EV), and New Hypotheses (NH).}
\label{fig:Phyto_indic_associations}
\end{figure}

The top hundred indication-phytochemical relationships obtained based on specificity scores are provided in \href{https://docs.google.com/spreadsheets/d/15dokKTu2JViNgCvz7f66ugYKPg-wU18TZe1eVI3OCeA/edit?gid=0#gid=0}{Table S1}. A bipartite representation of these relationships is shown in Fig.~\ref{fig:Phyto_indic_associations}.

\subsection{Minimum Set Cover of Recipes and Disease Categories}
\begin{figure}[!h]
\centering
   \centering
   \includegraphics[width=\textwidth]{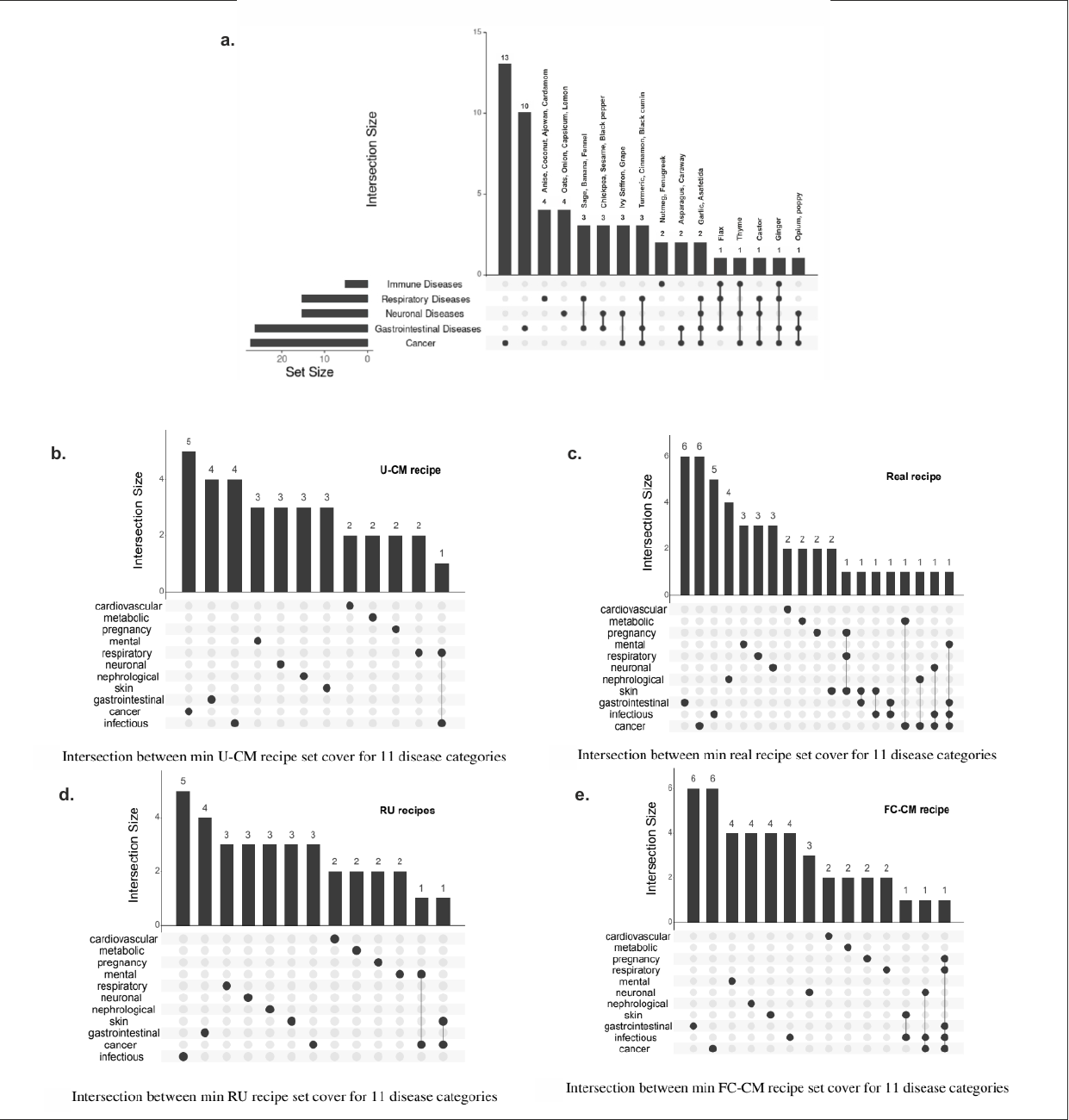} 
   \caption{ Combined minimal set showing the smallest number of phytochemicals needed to cover all disease categories across various recipe categories. These results aid in understanding the distribution and representation of diseases within different recipe datasets and optimizing dietary recommendations.}
   \label{fig:Min_set_cover}
\end{figure}

In Fig.~\ref{fig:Min_set_cover}, the minimum set cover plot helps us understand which recipes cover the most disease categories with the fewest ingredients. It shows intersections of recipe sets for different disease categories. In Fig.~\ref{fig:Min_set_cover}a, we see the overall intersection size of various disease categories, indicating which health issues are most frequently addressed by the recipes. Fig.~\ref{fig:Min_set_cover}b-e breaks this down further for different recipe types: U-CM, Real, RU, and FC-CM recipes. 
   
   For each recipe type, the plot shows how many disease categories each recipe can help with, highlighting the overlapping health benefits. The bars represent the number of shared disease categories between recipe sets. This helps identify recipes that are most versatile and beneficial across multiple health issues, providing insights into the potential wide-ranging impact of different phytochemical recipes.

\subsection{Gene-Phytochemical Interactions}

\begin{figure}[H]
\centering
   \centering
   \includegraphics[width=\textwidth]{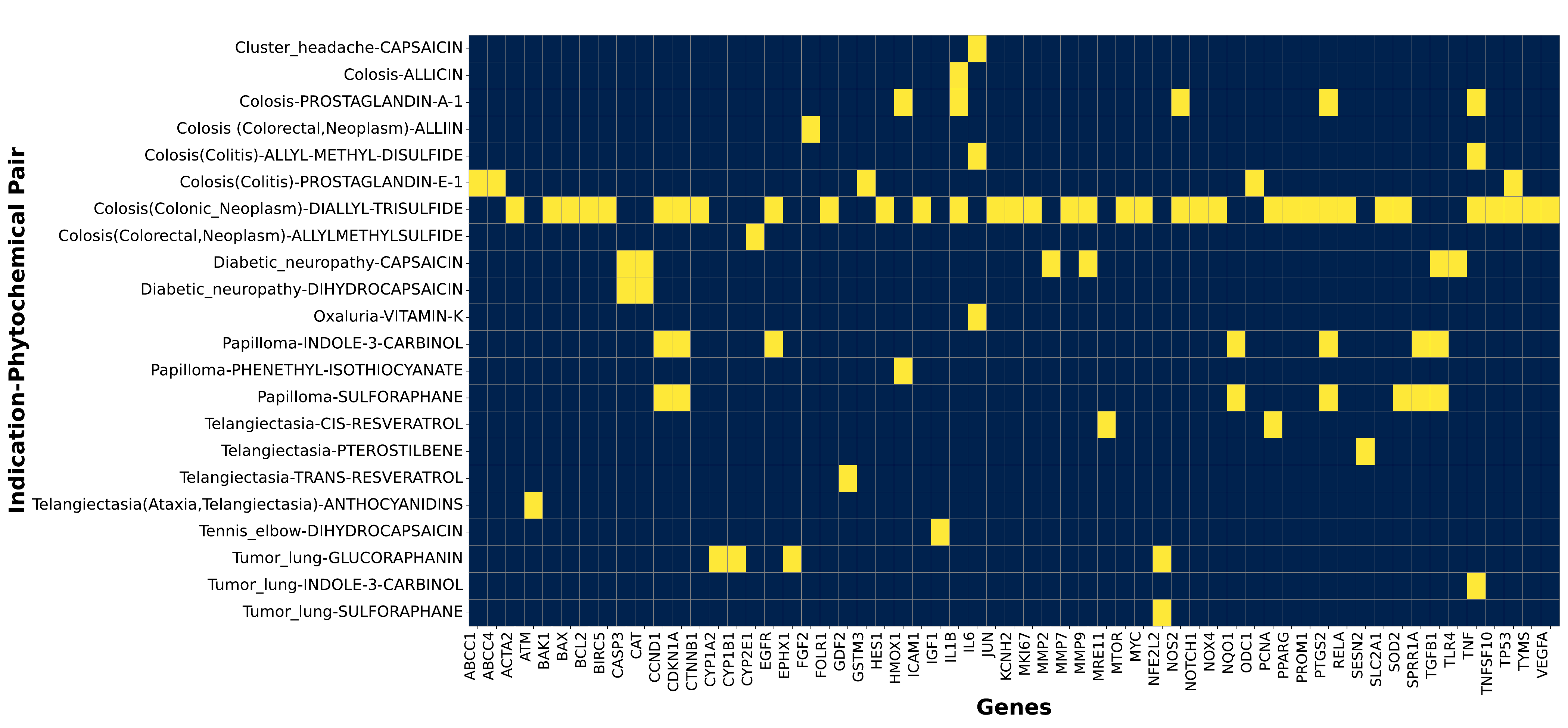} 
   \caption{Heat map illustrating various phytochemicals, affecting specific indications that interact with different genes. Here, the vertical axis represents combinations of health indications and phytochemicals, whereas the horizontal axis lists the specific genes. Yellow cells indicate the presence of a gene interaction for a given phytochemical-indication combination.}
   \label{fig:gene_heatmap}
\end{figure}

Using the Toxicogenomics Database (CTD) as the source, we inferred interactions between genes and phytochemicals.
The specificity score enabled us to infer that capsaicin and its three derivatives (dihydrocapsaicin, capsidiol, and capsanthin), found in hot peppers and capsicum, have therapeutic benefits for diseases such as tennis elbow, cluster headaches, and diabetic neuropathy. 
We also found that sulforaphane and indole-3-carbinol, found in kale, kohlrabi, etc. mitigate symptoms of lung tumors and related disorders interacting with genes like NFE2L2, NQO1, HMOX1, etc. \cite{Zhang2019}, with a CTD score of 52.85 (see Fig.~\ref{fig:gene_heatmap}). These findings are well-supported by CTD and literature \cite{Zhang2019, Choi2018}. Specifically, CTD inferred that dihydrocapsaicin and diabetic neuropathies are related due to their shared association with the CASP3 and CAT genes. Literature has shown that capsaicin and its derivatives have been used for treating a wide range of pain problems, including diabetic neuropathy, postoperative pain, and cluster headaches \cite{Basith2016, Peppin2014, Choi2018}.
 CTD made the same inferences, linking indole-3-carbinol to lung tumors through genes such as TNF, CYP2A6, EGFR, IRF1, PRKN, and IL1B, and to papillomas through genes such as TGFB1, PTGS2, EGFR, SPRR1A, and NQO1.
 Thus, our systematic validation demonstrates the effectiveness of the indication-phytochemical bipartite graph in uncovering new associations and confirming known relationships between phytochemicals and diseases. The high validation rate and the identification of previously unknown relationships with strong supporting evidence exhibit the potential of this approach for finding disease-phytochemical relationships.

\end{document}